\documentclass{article}
\usepackage{dcase2016,amsmath,graphicx,url,times}
\usepackage{color,soul} 
\usepackage{multirow}

\title{SOUND EVENT DETECTION IN MULTICHANNEL AUDIO USING SPATIAL AND HARMONIC FEATURES}

\name{Sharath Adavanne, Giambattista Parascandolo, Pasi Pertil\"{a}, Toni Heittola, Tuomas Virtanen\thanks{The research leading to these results has received funding from the European Research Council under the European Union’s H2020 Framework Programme through ERC Grant Agreement 637422 EVERYSOUND, and Google Faculty Research Award project ``Acoustic Event Detection and Classification Using Deep Recurrent Neural Networks". The authors also wish to acknowledge CSC-IT Center for Science, Finland, for computational resources.}}
\address{Department of Signal Processing, Tampere University of Technology}

\begin{document}

\ninept
\maketitle

\begin{sloppy}

\begin{abstract}
In this paper, we propose the use of spatial and harmonic features in combination with long short term memory (LSTM) recurrent neural network (RNN) for automatic sound event detection (SED) task. Real life sound recordings typically have many overlapping sound events, making it hard to recognize with just mono channel audio. Human listeners have been successfully recognizing the mixture of overlapping sound events using pitch cues and exploiting the stereo (multichannel) audio signal available at their ears to spatially localize these events. Traditionally SED systems have only been using mono channel audio, motivated by the human listener we propose to extend them to use multichannel audio. The proposed SED system is compared against the state of the art mono channel method on the development subset of TUT sound events detection 2016 database \cite{dcase2016Data}. The usage of spatial and harmonic features are shown to improve the performance of SED.
\end{abstract}


\section{Introduction}
\label{sec:intro}
A sound event is a segment of audio that a human listener can consistently label and distinguish in an acoustic environment. The applications of such automatic sound event detection (SED) are numerous; embedded systems with listening capability can become more aware of its environment \cite{environmentalSED}\cite{robot}. Industrial and environmental surveillance systems and smart homes can start automatically detecting events of interest \cite{surveillance}. Automatic annotation of multimedia can enable better retrieval for content based query methods \cite{contentRetrieval}. 

The task of automatic SED is to recognize the sound events in a continuous audio signal. Sound event detection systems built so far can be broadly classified to monophonic and polyphonic. Monophonic  systems are trained to recognize the most dominant of the sound events in the audio signal \cite{monoPhonic2013}. While polyphonic systems go beyond the most dominant sound event and recognize all the overlapping sound events in a segment \cite{monoPhonic2013}\cite{polyHough2013}\cite{emre2015}\cite{giam2016}. We propose to tackle such polyphonic soundscape which replicates real life scenario in this paper. 

Some SED systems have tackled polyphonic detection using mel-frequency cepstral coefficients (MFCC) and hidden Markov models (HMMs) as classifiers with consecutive passes of the Viterbi algorithm \cite{Mesaros2010_EUSIPCO}. In \cite{nmf}, a non-negative matrix factorization was used as a pre-processing step, and the most prominent event in each of the stream was detected. However, it still had a hard constraint of estimating the number of overlapping events. This was overcome by using coupled NMF in \cite{coupledNmf}. Dennis et al \cite{polyHough2013} took an entirely different path from the traditional frame-based features by combining generalized Hough transform (GHT) with local spectral features. 

More recently, the state of the art SED systems have used log mel-band energy features in DNN \cite{emre2015}, and RNN-LSTM \cite{giam2016} networks trained for multi-label classification. Motivated by the good performance of RNN-LSTM over DNN as shown in \cite{giam2016}, we continue to use the multi-label RNN-LSTM network.

The present state of the art polyphonic SED systems have been using a single channel of audio for sound event detection. Polyphonic events can potentially be tackled better if we had multichannel data. Just like humans use their two ears (two channels) to recognize and localize the sound events around them \cite{iid_itd}, we can also potentially train machines to learn sound events from multichannel of audio. Recently, Xiao et al \cite{multiChannelASR} have successfully used spatial features from multichannel audio for far field automatic speech recognition (ASR) and shown considerable improvements over just using mono channel audio. This further motivates us to use spatial features for SED tasks. In this paper, we propose a spatial feature along with harmonic feature and prove its superiority over mono channel feature even with a small dataset of around 60 minutes. 

The remaining of the paper is structured as follows. We describe in Section 2 the features used and the proposed approach. Section 3 presents a short introduction to RNNs and long short-term memory (LSTM) blocks. Section 4 presents the experimental set-up and results on a database of real life recordings. Finally, we present our conclusions in Section 5.

\section{SOUND EVENT DETECTION}
\label{sec:sed}

The sound event detection task involves identifying temporally the locations of sound event and assigning them to one among the known set of labels. Sound events in real life have no fixed pattern. Different contexts, for example, forest, city, and home have a different variety of sound events. They can be of different sparsity based on the context, and can occur in isolation or be completely overlapped with other sound events. While recognizing isolated sounds have been done with an appreciable accuracy \cite{isolatedSED}, detecting the mixture of labels in an overlapped sound event is a challenging task, where still a considerable amount of improvements can be made. Figure \ref{fig:multiLabel} shows a snippet of sound event annotation, where three sound events - speech, car, and dog bark happen to occur. At time frame \textit{t}, two events - speech and car are overlapping. An ideal SED system should be able to handle such overlapping events. 

The human auditory system has been successfully exploiting the stereo (multichannel) audio information it receives at its ears to isolate, localize and classify the sound events. A similar set up is envisioned and implemented, where the sound event detection system gets a stereo input and suitable spatial features are implemented to localize and classify sound events.

The proposed sound event detection system, shown in Figure \ref{fig:framework}, works on real life multichannel audio recordings and aims at detecting and classifying isolated and overlapping sound events.

Three sets of features -log mel-band energies, pitch frequency, and its periodicity, and time difference of arrival (TDOA) in sub-bands, are extracted from the stereo audio. All features are extracted at a hop length of 20 ms to have consistency across features.

\begin{figure}
  \centering
  \includegraphics[width=\columnwidth]{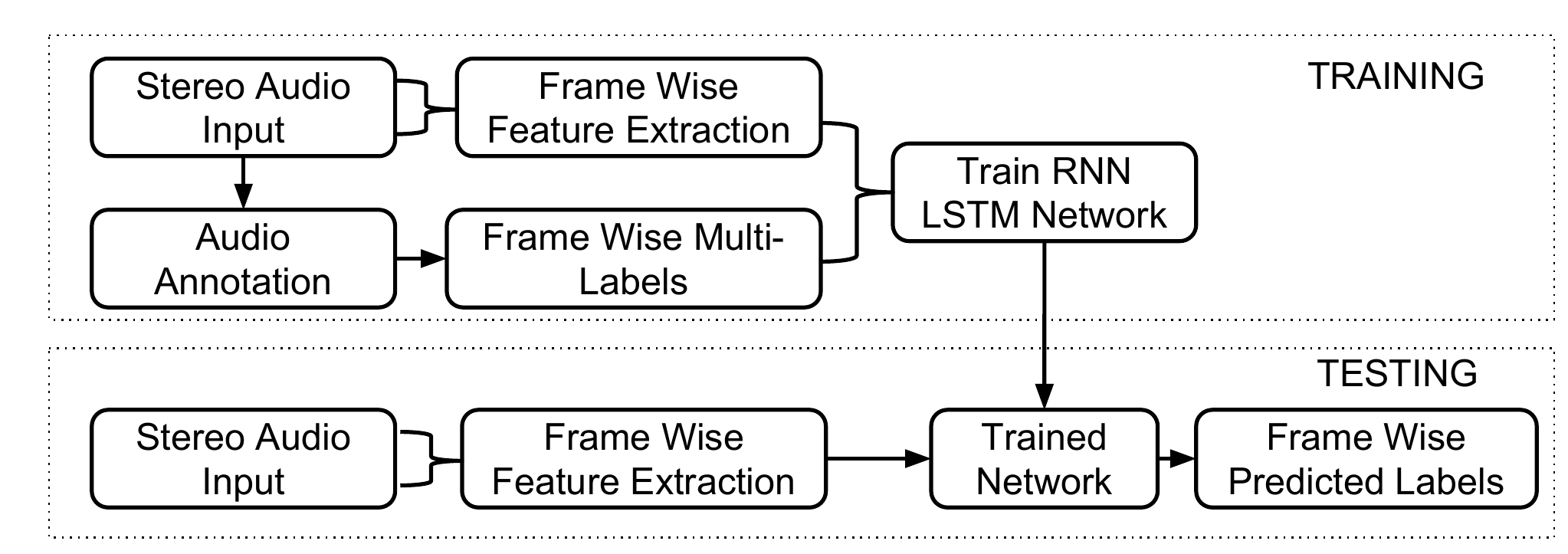}\vspace{-5pt}
  \caption{Framework of the training and testing procedure for the proposed system.}\vspace{-15pt}
  \label{fig:framework}
\end{figure}

\subsection{Log mel-band energy}
\label{ssec:mel}
Log mel-band energies have been used for mono channel sound event detection extensively \cite{emre2015}\cite{giam2016}\cite{sainath2015} and have proven to be good features. In the proposed system we continue to use log mel-band energies, and extract it for both the stereo channels. This is motivated from the idea that human auditory system exploits the interaural intensity difference (IID) for spatial localization of sound source \cite{iid_itd}. Neural networks are capable of performing linear operations, which includes the difference. Therefore, when trained on the stereo log mel-band energy data, it will learn to obtain information similar to IID.

Each channel of the audio is divided into 40 ms frames with 50\% overlap using hamming window. Log mel-band energies are then extracted for each of the frames ($mel$ in Table \ref{table:1}). We use 40 mel-bands spread across the entire spectrum.

\subsection{Harmonic features}
\label{ssec:pitch}
The pitch is an important perceptual feature of sound. Human listeners have evolved to identify different sounds using the pitch cues, and can make efficient use of pitch to acoustically separate each of the mixture in an overlapping sound event \cite{pitchSED1990}. Uzkent et al \cite{uzkent2012} have shown improvement in accuracy of non speech environmental sound detection used pitch range along with MFCC's. Here we propose using the absolute pitch and its periodicity as the features ($pitch$ in Table \ref{table:1}).

The librosa implementation of pitch tracking \cite{librosa} on thresholded parabolically-interpolated STFT \cite{jos} was used to estimate the pitch and periodicity.

Since we are handling multi-label classification it is intuitive to identify as many dominant fundamental frequencies as possible and use them to identify the sound events. The periodicity feature gives the confidence measure for the extracted pitch value and helps the classifier to make better decisions based on pitch.

The overlapping sound events in the training data (Section \ref{ssec:dataset}) did not have more than three events overlapping at a time, hence we have limited ourselves to using the top three dominant pitch values per frame. So, for each of the channels, top three pitch values, and its respective periodicity values are extracted at every frame in 100-4000 Hz frequency range ($pitch3$ in Table \ref{table:1}).

\begin{figure}
  \centering
  \includegraphics[width=\columnwidth]{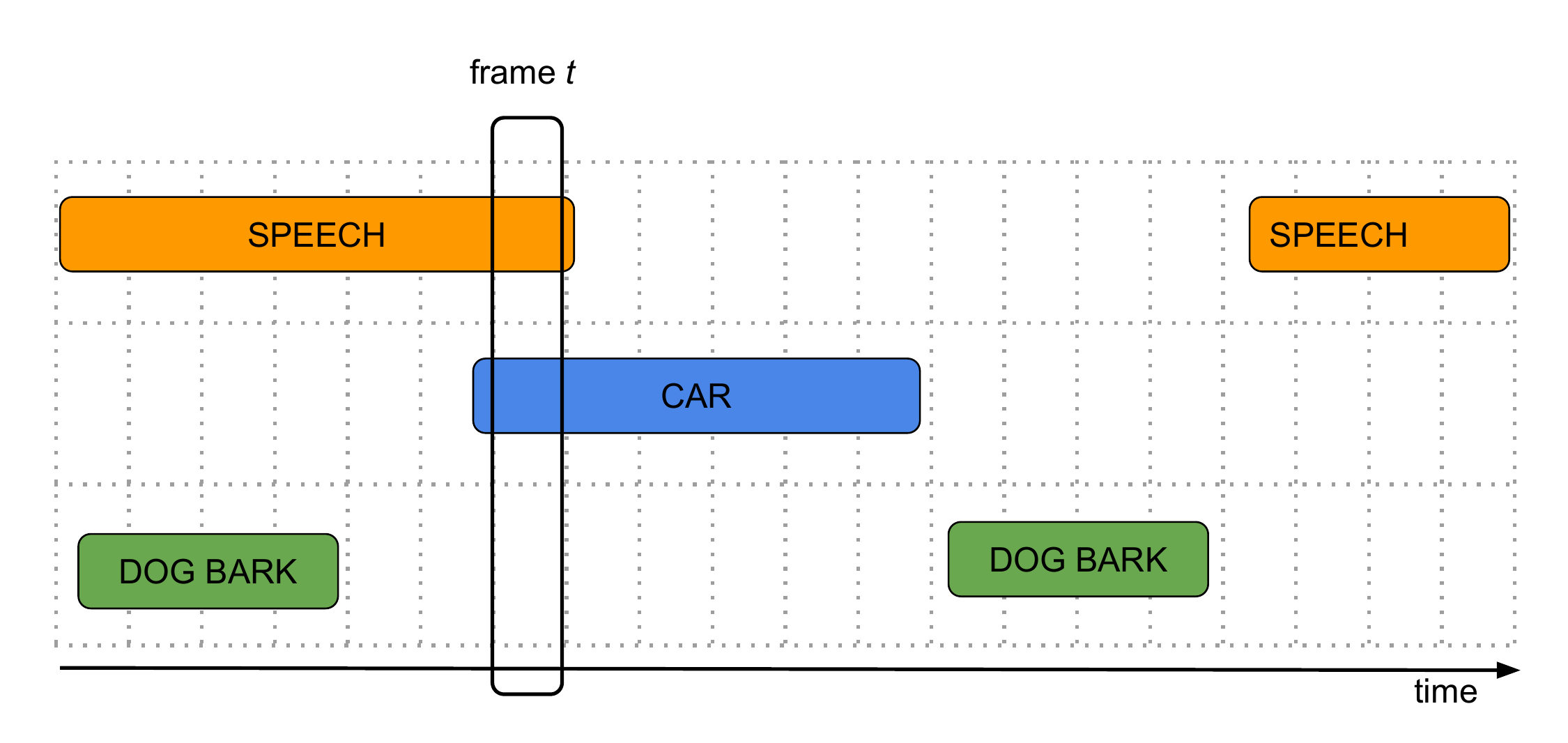}\vspace{-10pt}
  \caption{Sound events in a real life scenario can occur in isolation or overlapped. We see that at frame \textit{t}, speech and car events are overlapping.}
  \vspace{-15pt}
  \label{fig:multiLabel}
\end{figure}

\begin{table*}
\centering
\begin{tabular}{| c | c | p{10cm} |}
\hline
Feature Name & Length & Description\\
\hline
$mel$ & 40 & Log mel-band energy extracted on a single channel of audio\\ 
$pitch$ & 2 & Most dominant pitch value and periodicity extracted on a single channel \\
$pitch3$ & 6 & Top three dominant pitch and periodicity values extracted on a single channel\\
$tdoa$ & 5 &  Median of multi-window TDOA's extracted from stereo audio \\
$tdoa3$ & 15 & Concatenated multi-window TDOA's extracted from stereo audio \\
\hline
\end{tabular}
\caption{Definitions of acoustic features proposed for sound event detection.}\vspace{-15pt}
\label{table:1}
\end{table*}

\subsection{Time difference of arrival (TDOA) features}
\label{ssec:spatialFeat}
Overlapping sound events have forever troubled classification systems. This is mainly because the feature vector for the overlapped frame is a combination of different sound events. But, human listeners have been able to successfully identify each of the overlapping sound events by isolating and localizing the source spatially. This has been possible due to the interaural time delay (ITD) \cite{iid_itd}

Each sound event has its own frequency band, some occur in low frequencies, some in high, and some occur all across the frequency band. If we can divide the frequency spectrum into different bands, and identify the spatial location of the sound source in each of these bands, then this is an extra dimension of the feature, which the classifier can learn to estimate the number of possible sources in each frame, and their orientation in the space. We implement this by dividing the spectral frame into five mel-bands and calculating the time difference of arrival (TDOA) at each of these bands.

For example, if a non-overlapping isolated sound event is spread across the entire frequency range, and we are calculating the TDOA in five mel-bands. We should have the same TDOA values for each of the bands. However, if we have two overlapping sounds $S_1$ and $S_2$, where $S_1$ is spread in the first two bands and $S_2$ is spread in the last two bands. The feature vector will have different TDOA values for each of the sounds, which the classifier can learn to isolate and identify them as separate sound events.

The TDOA can be estimated using the generalized cross-correlation with phase-based weighting (GCC-PHAT)~\cite{Knapp_Carter-1976}. Here, we extract the correlation for each mel-band separately:
 \begin{equation}
 R_b(\Delta_{12},t) = \sum_{k=0}^{N-1} H_b(k) \frac{  X_1(k,t) \cdot X_2^\ast(k,t)}{\arrowvert X_1(k,t)\arrowvert \arrowvert X_2(k,t)  \arrowvert} e^{ i 2\pi k \Delta_{12} / N },
 \end{equation}
 where $N$ is the number of frequency bands, $X(k,t)$ is the FFT coefficient of the $k$th frequency band at time frame $t$ and the subscript specifies the channel number, ${H}_b(k)$ is the magnitude response of the $b$th mel-band of total of $B$ bands and $\Delta_{12}$ is the sample delay value between channels. The TDOA is extracted as the location of correlation peak magnitude for each mel-band and time frame.
\begin{equation}
 \tau(b,t)= \underset{\Delta_{12}}{\textrm{argmax}}\left\{ R_b(\Delta_{12},t)\right\}
\end{equation}

The maximum and minimum TDOA values are truncated between values 
 $-2\tau_\textrm{max},2\tau_\textrm{max}$, where 
 $\tau_\textrm{max}$ is the maximum sample delay between a sound wave traveling between microphones. 

The sound events in the training set were seen to be varying from 50 ms to a few seconds. In order to accommodate such variable length sound events, TDOA was calculated in three different window lengths --- 120, 240 and 480 ms, with a constant hop length of 20 ms. The TDOA values of these three windows were concatenated for each mel-band to form one set of TDOA features. So, TDOA values extracted in five mel-band, and for three window lengths, on concatenation gives 15 TDOA values per frame ($tdoa3$ in Table \ref{table:1}).

TDOA values in small windows are generally very noisy and unreliable. To overcome this, the median of the TDOA values from the above three different window lengths for each sub-band of the frame was used as the second set of TDOA features ($tdoa$ in Table \ref{table:1}). Post filtering across window lengths, the TDOA values in each mel-band were also median filtered temporally using a kernel of length three to remove outliers.

\section{MULTI-LABEL RECURRENT NEURAL NETWORK BASED SOUND EVENT DETECTION}
\label{sec:rnn}

Deep neural networks have shown to perform well on complex pattern recognition tasks, such as speech recognition~\cite{graves2013speech}, image recognition~\cite{krizhevsky2012imagenet} and machine translation~\cite{bahdanau2014neural}. A deep neural network typically computes a map from an input to an output space through several subsequent matrix multiplications and non-linear activation functions. The parameters of the model, i.e. its weights and biases, are iteratively adjusted using a form of optimization such as gradient descent.

When the network is a directed acyclic graph, i.e. information is only propagated forward, it is known as a feedforward neural network (FNN). When there are feedback connections the model is called a recurrent neural network (RNN). An RNN can incorporate information from previous timesteps in its hidden layers, thus providing context information for tasks based on sequential data, such as temporal context in audio tasks. Complex RNN architectures --- such as long short-term memory (LSTM) \cite{hochreiter1997long} --- have been proposed in recent years in order to attenuate the vanishing gradient problem \cite{bengio1994learning}. LSTM is currently the most widely used form of RNN, and the one used in this work as well.

In SED, RNNs can be used to predict probabilities for each class to be active in a given frame at timestep $t$. The input to the network is a sequence of feature vectors $\bf{x}(\textmd{t})$; the network computes hidden activations for each hidden layer, and at the output layer a vector of predictions for each class $\bf{y}(\textmd{t})$. A sigmoid activation function is used at the output layer in order to allow several classes to be predicted as active simultaneously. By thresholding the predictions at the output layer it is possible to obtain a binary activity matrix.

\subsection{Neural network configurations}
\label{ssec:nn}

 For each recording, we obtain a sequence of feature vectors, which is normalized to zero mean and unit variance, and the scaling parameters are saved for normalizing the test feature vectors. The sequences are further split into non-overlapping sequences of length 25 frames. Each of these frames has a target binary vector, indicating which classes are present in the feature vector.
 
 We use a multi-label RNN-LSTM with two hidden layers each having 32 LSTM units. The number of units in the input layer depends on the length of the feature being used. The output layer has one neuron for each class. The network is trained by back propagation through time (BPTT) \cite{bptt1990} using binary cross-entropy as loss function, Adam optimizer \cite{adamKeras} and block mixing \cite{giam2016} data augmentation. Early stopping is used to reduce over-fitting, the training is halted if the segment based error rate (ER) (see Section \ref{ssec:metrics}) on the validation set does not decrease for 100 epochs.
 
At test time we use scaling parameters estimated on training data to scale the feature vectors and present them in non-overlapping sequences of 25 frames, and threshold the outputs with a fixed threshold of 0.5, i.e., we mark an event is active if the posterior in the output layer of network is greater than 0.5 and otherwise inactive.

\renewcommand{\thefootnote}{\fnsymbol{footnote}}

\begin{table*}[t]
\centering
\begin{tabular}{| p{5cm} | p{3cm} | c | c | c | c | c | c |}
\hline
& Feature combination  & \multicolumn{2}{|c|}{Home} & \multicolumn{2}{|c|}{Residential area} & \multicolumn{2}{|c|}{ Average} \\ 
\cline{3-8}
& & ER & F (\%) & ER & F (\%) & ER & F (\%)  \\
\hline

Baseline system using GMM classifier in DCASE 2016 \cite{dcase2016Data}\cite{dcase2016task3web} & $mfcc;delta;acc$ & 0.96 & 15.9 & 0.86	& 31.5 & 0.91 &23.7 \\
\hline

Mono channel feature With RNN-LSTM network & $mel_1$ & \bf0.94 & \bf27.4 & 0.88 & 38.3 & 0.91 & 32.9\\ 
\hline

\multirow{4}{5cm}{Hybrid (mono and stereo) features with RNN-LSTM network}& $mel_1;pitch_1$ & 0.97 & 25.4 & 0.85 & 43.4 & 0.91 & 34.4\\ 
\multirow{4}{*}{}&  $mel_1;pitch3_1$ & 0.96 & 27.6 & 0.88 & 43.9 & 0.92 & 35.7\\ 
\multirow{4}{*}{}& $mel_1;tdoa$ & 1.02 & 19.4 & 0.89 & 40.2 & 0.96 & 29.8\\ 
\multirow{4}{*}{}& $mel_1;tdoa3$ & 0.98 & 25.9 & 0.87 & 40.5 & 0.92 & 33.2\\ 
\hline

\multirow{9}{5cm}{Stereo features with RNN-LSTM network}&   $mel_2$  & 1.03 & 25.4 & 0.84 & 45.9 & 0.93 & 35.6\\ 
\multirow{9}{*}{}&    $mel_2;pitch_2$ & 1.03 & 24.9 & 0.93 & 40.9 & 0.98 & 32.9\\ 
\multirow{9}{*}{}&    $mel_2;pitch3_2$ & 0.97 & 26.6 & 0.88 & 41.7 & 0.92 & 34.2\\ 
\multirow{9}{*}{}&    $mel_2;tdoa$ & 1.01 & 24.4 & \bf0.82 & \bf46.4 & \bf0.91 & \bf35.4\\ 
\multirow{9}{*}{}&    $mel_2;tdoa3$ & 0.96 & 24.9 & 0.86 & 38.5 & 0.91 & 31.7\\ 
\multirow{9}{*}{}&    $mel_2;tdoa3;pitch_2$ & 0.97 & 25.7 & 0.85 & 43.1 & 0.91 & 34.4\\ 
\multirow{9}{*}{}&    $mel_2;tdoa3;pitch3_2$ & 0.99 & 26.5 & 0.91 & 35.2 & 0.95 & 30.9\\ 
\multirow{9}{*}{}&    $mel_2;tdoa;pitch_2$ & 0.98 & 24.7 & 0.87 & 43.8 & 0.92 & 34.2\\ 
\multirow{9}{*}{}&    $mel_2;tdoa;pitch3_2$ & 0.94 & 26.3 & 0.89 & 40.5 & 0.91 & 33.4\\ 
\hline
\end{tabular}
\caption{ Segment based error rate (ER) and F-score achieved for different feature combinations in home and residential area contexts for the development set. The features listed in Table \ref{table:1} are used in different combinations with the proposed RNN-LSTM network. The subscripts '$_1$' and '$_2$' in the feature combinations column represent how many channels the features were extracted on. For example, feature combination \textit{$mel_2;tdoa;pitch_2$} means that the final feature vector has log mel-band energies, most dominant pitch and periodicity values extracted on both the stereo channels, and the time difference of arrival (TDOA) calculated between the stereo channels. The highlighted ER and F-score pair for each context is the best ER score achieved.}\vspace{-15pt}
\label{table:2}
\end{table*}

\section{EVALUATION AND RESULTS}
\label{sec:eval}

\subsection{Dataset}
\label{ssec:dataset}

We evaluate the proposed SED system on the development subset of TUT sound events detection 2016 database \cite{dcase2016Data}. This database has stereo recordings which were collected using binaural Soundman OKM II Klassik/studio A3 electret in-ear microphones and Roland Edirol R09 wave recorder using 44.1 kHz sampling rate and 24-bit resolution. It contains two contexts - home and residential area. Home context has 10 recordings with 11 sound event classes and the residential area context has 12 recordings with 7 classes. The length of these recordings is between 3-5 minutes.

In the development subset provided, each of the context data is already partitioned into four folds of training and test data. The test data was collected such that each recording is used exactly once as the test, and the classes in it are always a subset of the classes in the training data. Also, 20\% of the training data recordings in each fold were selected randomly to be used as validation data. The same validation data was used across all our evaluations.

\subsection{Metrics}
\label{ssec:metrics}
We perform the evaluation of our system in a similar fashion as \cite{dcase2016Data} which uses the established metrics for sound event detection defined in \cite{dcase2016Metrics}. The error rate (ER) and F-scores are calculated on one second long segments. The results from all the folds are combined to produce a single evaluation. This is done to avoid biases caused due to data imbalance between folds as discussed in \cite{apple}.

\subsection{Results}
\label{ssec:results}
The baseline system for the dataset \cite{dcase2016Data} uses 20 static (excluding the 0th coefficient), 20 delta and 20 acceleration MFCC coefficients extracted on mono audio with 40 ms frames and 20 ms hop length. A Gaussian mixture model (GMM) consisting of 16 Gaussians is then trained for each of the positive and negative values of the class. This baseline system gives a context average ER of 0.91 and F-score of 23.\%. An ideal system should have an ER of 0 and an F-score of 100\%. 

In Table \ref{table:2} we compare the segment based ER and F-score for different combinations of proposed spatial and harmonic features. In all these evaluations, only the size of the input layer changes based on the feature set, with the rest of the configurations in the RNN-LSTM network remaining unchanged. 

Mono channel audio was created by averaging the stereo channels in order to compare the performance of the proposed spatial and harmonic features for multichannel audio. One of the present state of the art SED system for mono channel is proposed in \cite{giam2016}. An RNN-LSTM network is trained in a similar fashion with log mel-band energy feature (Section \ref{ssec:mel}) and evaluated. Across contexts, the F-score was seen to be better than the GMM baseline system with comparable ER. Here onwards we use this mono-channel log mel-band feature and RNN-LSTM network configuration result as a baseline for comparisons.

A set of hybrid combinations were tried as shown in Table \ref{table:2}. All combinations other than $mel_1;tdoa$ performed better than the baseline across contexts in F-score.

Finally, the full spectrum of proposed spatial and harmonic features were evaluated in different combinations with RNN-LSTM network. With a couple of exceptions - $mel_2;pitch_2$ and  $mel_2;tdoa3;pitch3_2$, all the combinations of features performed equal to or better than the baseline in average F-scores, with marginally similar average ER as baseline. Given the dataset size of around 60 minutes, it is difficult to conclusively say that the binaural features are far superior to monaural features; but they surely look promising. 

Binaural features - $mel_2$ and $mel_2;tdoa;pitch_2$ in Table \ref{table:3} were submitted to the DCASE 2016 challenge \cite{dcase2016task3web},  where they were evaluated as the top performing systems. Monaural feature $mel_1$ was submitted unofficially to compare the performance with binaural features. The hyper-parameters of the network were tuned before the submission, and hence the development set results in Table \ref{table:3} are different from Table \ref{table:2}. Three hidden layers with 16 LSTM units each were used for $mel_2$, while  $mel_1$ and $mel_2;tdoa;pitch_2$ were trained with two layers each having 16 LSTM units.

\begin{table}[h]
\centering
\begin{tabular}{|l|c|c|c|c|}
\hline
\multicolumn{1}{|c|}{\multirow{2}{*}{\begin{tabular}[c]{@{}c@{}}Feature \\ combination\end{tabular}}} & \multicolumn{2}{c|}{\begin{tabular}[c]{@{}c@{}}Evaluation\\  dataset\end{tabular}} & \multicolumn{2}{c|}{\begin{tabular}[c]{@{}c@{}}Development\\ dataset\end{tabular}} \\ \cline{2-5} 
\multicolumn{1}{|c|}{} & ER & F (\%) & ER & F (\%) \\ 
\hline
$mel_1$ & 0.79 & 46.6 & 0.90 & 35.3 \\ \hline
$mel_2$  & 0.80 & 47.8 & 0.88 & 34.7 \\ \hline
$mel_2;tdoa;pitch_2$  & 0.88 & 37.9 & 0.87 & 34.8 \\ \hline
\end{tabular}
\caption{Comparison of segment based error rate (ER) and F-score for development and evaluation dataset. The evaluation dataset scores are the result of DCASE 2016 challenge \cite{dcase2016task3web}.} \vspace{-15pt}
\label{table:3}
\end{table}

\section{CONCLUSION}
\label{sec:conclusion}
In this paper, we proposed to use spatial and harmonic features for multi-label sound event detection along with RNN-LSTM networks. The evaluation was done on a limited dataset size of 60 mins, which included four cross validation data for two contexts --- home and residential area. The proposed multi-channel features were seen to be performing substantially better than the baseline system using mono-channel features. 

Future work will concentrate on finding novel data augmentation techniques. Augmenting spatial features is an unexplored space, and will be a challenge worth looking into. Concerning the model, further studies can be done on different configurations of RNN like extending them to bidirectional RNN's and coupling with convolutional neural networks.


\bibliographystyle{IEEEtran}
\bibliography{refs}

\end{sloppy}
\end{document}